# Evidence of spatial inhomogeneity near the onset of magnetically induced insulating state in superconducting thin films

C. L. Vicente, Y. Qin, and Jongsoo Yoon
*Department of Physics, University of Virginia, Charlottesville, VA 22903, U.S.A.*

Non-monotonic differential resistance ($dV/dI$) is observed in magnetically induced insulating films which exhibit apparent superconductor-metal-insulator transitions in the low temperature limit; at low bias currents the nonlinear transport is insulator-like while at high bias currents it is characteristic of metallic phase. The non-monotonic $dV/dI$ may be evidence that the insulating state consists of metallic domains connected by point contacts (insulating gaps), implying that spatial inhomogeneities play a dominant role in determining the nature of the apparent metal-insulator transition.

The interplay of disorder, Coulomb interaction, and the superconducting pairing mechanism is particularly interesting in two-dimensions (2D) since 2D is the lower critical dimension for both localization and superconductivity. Traditionally, it has been believed that in 2D the disorder- or magnetic field-induced suppression of superconductivity leads to a direct superconductor-insulator transition (SIT) in the limit of zero temperature ($T \to 0$) [1-5]. Many experimental results on the SIT have been analyzed in view of the so called "dirty boson" model [4 5], where the superconducting phase is described as a condensate of Cooper pairs with localized vortices and the insulating phase corresponds to a Bose glass state which is a condensate of vortices with localized Cooper pairs. On the other hand, many authors [6-9] have emphasized that spatial inhomogeneity plays a central role in determining the nature of the SIT. It has been predicted that mechanisms such as localized gapless electronic excitations [10] critically enhanced mesoscopic fluctuations [8], or inherent random nature of disorder [7,11] can cause significant local variations in the preference of one phase over the other. It has been argued that the SIT occurs via the resulting heterogeneous state where superconducting puddles form a percolating network in the background of insulating region. The percolation description for the nature of the SIT is an alternative to that by the dirty boson model where the SIT is understood in terms of Cooper pair scattering out of superconducting condensate into a Bose glass state.

While the nature of the SIT is still an open issue, the interest on the problem has been further intensified as an apparent metallic behavior in the low temperature limit is observed, notably in amorphous MoGe [12-14] and Ta [15,16] thin films under weak magnetic fields ($B$). The metallic behavior is characterized by a drop in resistance ($\rho$) followed by a saturation to a finite value with approaching the zero temperature limit. The occurrence of the metallic behavior over a significant range of magnetic fields indicates the possibility of an unexpected metallic phase intervening the superconducting and insulating phase. While Ng and Lee [17], and Tewari [18] described the metallic behavior as a crossover at finite temperatures and argued that there should be no metallic phase at $T = 0$, others predicted an emergence of true metallic ground state when the superconductivity is suppressed. Galitski et al. [19] suggested that the vortex metal phase derived by treating vortices as fermions can exhibit finite resistance at $T = 0$. Das and Doniach [20] and Dalidovich and Phillips [21,22] proposed that the metallic phase might correspond to the Bose metal phase in which Cooper pairs lack phase coherence.

In this paper we consider the possible development of spatial inhomogeneity near the apparent $B$-induced metal-insulator transition in a nominally homogeneous system. Magnetic fields are applied perpendicular to the sample plane. The possible role of spatial inhomogeneity near the superconductor-metal transitions has been alluded to in the model of Spivak et al. [23] They have shown that a 2D system of superconducting grains imbedded in normal metal can exhibit a zero temperature superconductor-metal phase transition at arbitrarily large conductance of the system. We have studied Ta thin films which exhibit the unexpected metallic behavior in the limit of zero temperature [15]. Our measurements of the differential resistance ($dV/dI$) show that this quantity becomes non-monotonic when the magnetic field is increased to above the metal-insulator "critical" field $B_c$. At low current regime the nonlinear transport is characterized by a negative $d^2V/dI^2$ whereas at high current regime $d^2V/dI^2$ is positive. Recently, it has been shown that the nonlinear transport with a negative $d^2V/dI^2$ is unique to the insulating phase, and that a positive $d^2V/dI^2$ is a signature of the metallic mechanism [15,16]. We propose that the observed non-monotonic $dV/dI$ in the insulating phase near $B_c$ may be evidence of a spatially inhomogeneous state, where metallic domains are connected by point contacts (narrow gaps of insulating region) forming a percolating network. In this picture, at low bias currents the transport is dominated by the point contacts as they act as bottlenecks for the conduction of the metallic network. With increasing bias current, the point contacts become less resistive as manifested by their negative $d^2V/dI^2$, and at a sufficiently high bias current they no longer act as bottlenecks. As the transport of the network is dominated by the metallic domains at high current regime, the nonlinear transport exhibits a positive $d^2V/dI^2$; thus non-monotonic $dV/dI$. Our picture, which does not depend on the detailed mechanism of the metallic phase which is yet to be identified, suggests that the spatial inhomogeneity may play a dominant role in determining the nature of the apparent metal-insulator transition.

**Table 1.** List of sample parameters: nominal film thickness, mean field $T_c$ at $B = 0$, normal state resistivity at 4.2 K, critical magnetic field $B_c^*$ at which the resistance reaches 90 % of the high magnetic field saturation value, and correlation length calculated from $\xi = \sqrt{\Phi_o / 2\pi B_c^*}$ where $\Phi_o$ is the flux quantum.

| samples | t (nm) | $T_c$ (K) | $\rho_n$ (Ω/□) | $B_c^*$ (T) | ξ (nm) |
|---|---|---|---|---|---|
| 1 | 5.0 | 0.675 | 1180 | 0.88 | 19 |
| 2 | 4.5 | 0.622 | 1100 | 1.03 | 18 |
| 3 | 5.5 | 0.625 | 840 | 1.07 | 18 |

Our samples are produced by dc sputtering Ta on Si substrates. Samples are patterned into a bridge, 1 mm wide and 5 mm long, for the standard four point measurements with a shadow mask. Prior to the deposition, the sputter chamber is baked at ~ 110 °C for several days reaching a base pressure of ~ $10^{-8}$ Torr. The chamber and Ta source were cleaned by presputtering for ~ 30 min at a rate of ~ 1 nm/s. Films are grown at a rate of ~ 0.05 nm/s at an Ar pressure of ~ 4 mTorr. The superconducting properties of our Ta thin films at $B = 0$ are characteristic of homogeneously disordered thin films [24], and consistent with the results of x-ray structural investigations [15]. We have studied more than a dozen samples with their mean field superconducting transition temperatures ($T_c$) in the range 0.2 – 0.7 K. All the films have shown consistent results. In this paper we describe data from 3 samples whose parameters are summarized in Table 1.

The differential resistance measured at 60 mK across the magnetic field driven metal-insulator transition for a 5.0 nm thick tantalum film is shown in Fig. 1(a). At magnetic fields 0.9 T or below (dashed lines), the $dV/dI$ is a monotonically increasing function of bias current, which has been established as characteristic of the metallic phase [15, 16]. Shown in Fig. 1(b) is the temperature dependence of the resistance in the field range 0 – 3 T. The metallic behavior is observed in the field 0.3 – 0.9 T. The fields 0.2 T or below correspond to the "superconducting" phase where the resistance at the lowest temperature is "immeasurably" small [bottom three traces in Fig. 1(b)].

When the magnetic field is increased to above 0.9 T, the $dV/dI$ becomes non-monotonic. The non-monotonic $dV/dI$, which is the main feature of this paper, is shown by two thick solid lines in Fig. 1(a) for $B = 0.95$ T and 1.00 T. At high bias current regime, $I > I_s$ where $I_s$ is the current for minima in $dV/dI$ and indicated by arrows, the sign of $d^2V/dI^2$ is positive as in the metallic phase of lower fields, indicating that the transport is dominated by the metallic mechanism. However, at low bias currents ($I < I_s$) the sign of $d^2V/dI^2$ is negative, which is the insulating characteristics [15, 16]. That the transport at low currents is insulating, is confirmed by an independent determination of the metal-insulator critical field $B_c$. Shown in Fig. 1(c) is the magnetoresistance measured in the low current limit (1 nA) at three different temperatures. In this plot, the critical field $B_c$ appears as a crossing point marked by an arrow at 0.908 T because of positive $d\rho/dT$ below $B_c$, negative $d\rho/dT$ above $B_c$, and $T$-independent $\rho$ at $B_c$.

The switching of the dominant transport mechanism from insulating at low currents to metallic at high currents could be understood in a simple picture for a heterogeneous state. In our picture, at least near $B_c$, the metallic and insulating regions coexist within the sample. At $B \lesssim B_c$, the metallic regions form a percolating network in the background of the insulating region. The metallic network is continuous and provides preferential current flow paths. Thus, the transport is dominated by the metallic mechanism at all bias currents. As the magnetic field is increased the metallic regions shrink and the insulating regions grow. When $B$ is increased to slight above $B_c$ ($B \gtrsim B_c$), the shrinkage of the metallic regions causes the metallic network to be broken into isolated domains or segments that are connected via point contacts which are narrow gaps of insulating region. The transport of the network at $B \gtrsim B_c$ in the low current regime is dominated by the point contacts because they act as

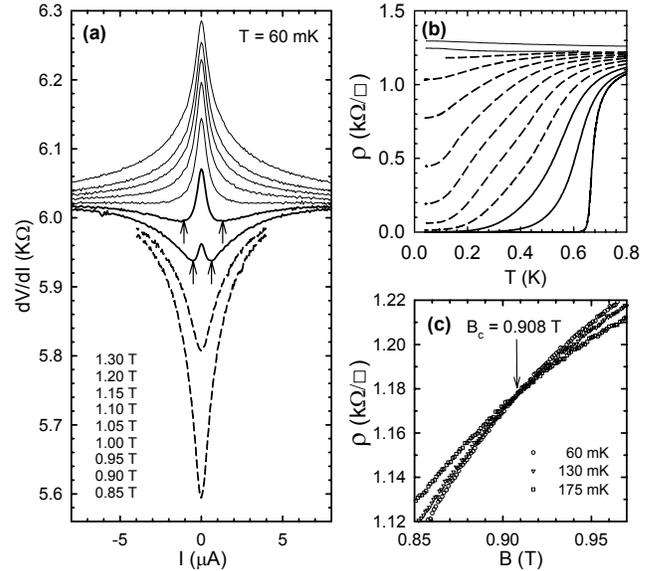

**FIG. 1** (a) $dV/dI$ of sample 1 at the indicated $B$ and $T$. The dashed lines correspond to the metallic phase and solid lines (both thick and thin) to the insulating phase when probed in the small current limit. The arrows indicate the current $I_s$ where $dV/dI$ exhibits minima. (b) Resistiviy of the same sample measured with a dc current of 1 nA at magnetic fields 0 – 1.0 T with 0.1 T interval and 3.0 T. The solid lines are for superconducting phase in the low $T$ limit, the dashed lines for the metallic phase, and the thin lines for the insulating phase. (c) Magnetoresistance measured with a dc current of 1 nA at the indicated $T$. The crossing point marked by the arrow corresponds to the metal-insulator critical field $B_c$.

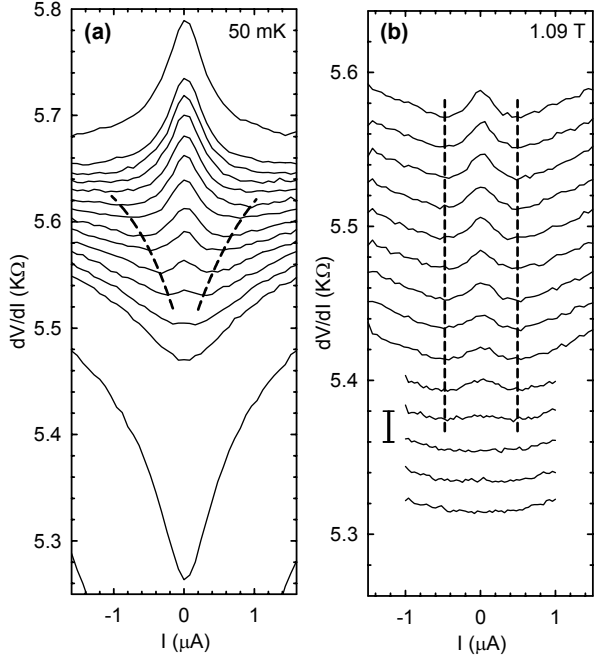
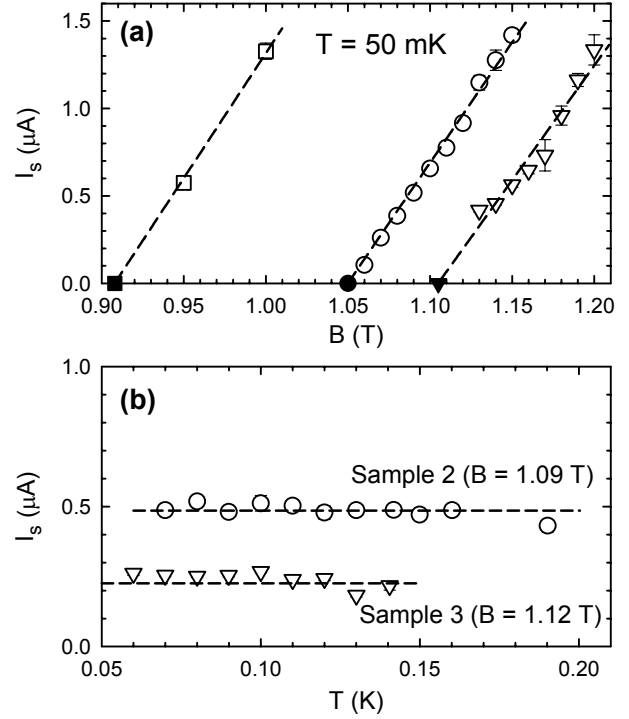

**FIG. 2** (a) $dV/dI$ of sample 2 at $B$, from the bottom, 0.90, 1.00, 1.05 − 1.16 T with a 0.01 T interval, and 1.20 T. The bias current was ac modulated with an amplitude of 1.5 nA at ~ 7 Hz. The dashed lines are to guide an eye to follow the minima in each trace. (b) $dV/dI$ of the same sample at $T$, from the top, 0.07 − 0.16 K with a 0.01 K interval, 0.19, 0.20, 0.22, and 0.25 K. Each trace is successively shifted by the amount indicated by the vertical scale bar. The dashed lines are to indicate minima in each trace.

**FIG. 3** (a) Square, circle, and triangle symbols are for sample 1, 2, and 3, respectively. Open symbols are for the current $I_s$ where minima in $dV/dI$ are found. The values of $I_s$ are determined by quadratic fittings of the data near each minimum. Filled symbols are the metal-insulator critical field of each sample obtained from the crossing point of magnetoresistance traces at several different temperatures. Dashed lines are to guide an eye. (b) Temperature-independence of $I_s$ is shown for the two samples.

bottlenecks for the transport. Therefore, the transport at $B \gtrsim B_c$ at low currents exhibits insulating characteristics.

At $B \gtrsim B_c$, at a sufficiently high bias current the transport properties can cross over from insulting to metallic because of the contrasting nonlinear transport in the metallic and insulating phase. The nonlinear transport of the insulating phase is characterized by a negative $d^2V/dI^2$, which means that the dynamic resistance ($V/I$) decreases with increasing current. The dynamic resistance of the metallic phase increases with increasing current as evidenced by positive $d^2V/dI^2$. Therefore, as $I$ increases the point contacts become less resistive while the metallic domains become more resistive. At a sufficiently high bias current, the point contacts no longer act as bottlenecks for transport; the transport in the network is then dominated by the metallic domains, resulting in a switch from insulating to metallic transport.

The assumption that the metallic domain shrinks and the insulating region grows with increasing $B$ has two inevitable consequences in terms of $B$-dependence of the characteristic current $I_s$. First, the current $I_s$ is expected to increase with increasing $B$. This is because the insulating gaps (point contacts) in the network become wider as $B$ increases. For a wider insulating gap, a larger bias current would be needed for the conductance of the gap to be large enough to cause the switching of the dominant transport mechanism. Second, the non-monotonic $dV/dI$ is likely to occur over a limited range of magnetic fields. This is because at sufficiently high magnetic fields where the insulating gaps become very wide the dominant transport is insulating at all bias current.

We observe both of the expected consequences. The increase of $I_s$ with $B$ is evident in the traces for 0.95 T and 1.00 T in Fig. 1(a). More detailed evolution of the $dV/dI$ minima is shown in Fig. 2(a) for another film. The dashed lines in Fig. 2(a) are to indicate the minima at $I_s$ for each trace. The current $I_s$ is found to almost linearly increase from zero with increasing $B$ above $B_c$, $I_s \propto (B - B_c)$. This is shown in Fig. 3(a). The values of $I_s$ in Fig. 3 are determined by fitting the data near each minimum to a quadratic form, $dV/dI = R_o + R_1 (I - I_s)^2$ where $R_o$, $R_1$, and $I_s$ are the fitting parameters. The filled symbols in Fig. 3(a) are the metal-insulator critical fields $B_c$ which are obtained from independent crossing point measurements like the one shown in Fig. 1(c). The non-monotonic $dV/dI$ is also found to occur only within about 10 − 15 % of $B_c$. This behavior was found in all (more than a dozen) the films we investigated. With increasing $B$, the minima in $dV/dI$ become shallower, disappear at $(B - B_c) / B_c \approx 0.1 - 0.15$, and at higher fields $dV/dI$ becomes monotonically decreasing function of increasing $I$. This trend is also clearly visible in the data set shown Fig. 1(a). The trace for 1.05 T in Fig, 1(a) does not show minima, and $d^2V/dI^2$ is "zero" at $I \gtrsim 2$ μA.

Interestingly, the current $I_s$ is found to be almost independent of temperature. Shown in Fig. 2(b) is an example of the $T$-driven evolution of a $dV/dI$ trace, which is non-monotonic at low temperatures. Note that each trace is successively shifted vertically. As temperature increases, the magnitude of the insulating feature (the low current portion where $d^2V/dI^2 < 0$) becomes progressively smaller and eventually disappears, whereas the values of $I_s$ remain almost independent of temperature. The temperature independent $I_s$ is emphasized by the dashed lines in Fig. 2(b) which indicate the minima in each $dV/dI$ trace. Figure 3(b) shows the values of $I_s$ as a function of temperatures, obtained by quadratic fittings, for two samples.

Within our simple picture, increasing $T$ could have two main effects. One is weakening of the strength of both the insulating and metallic mechanism as represented by the decrease in resistivity in the insulating and the increase in resistivity in the metallic regions. The other is the likely softening of the boundary between the two regions. The latter may induce effective changes in the pattern of the metallic network influencing the transport properties of the system. At present, it is not clear whether or how these two effects lead to the $T$-independent $I_s$. Quantitative understanding of the linear increase of $I_s$ with $B$ and $T$-independent $I_s$ requires thorough theoretical consideration of the problem.

It should be pointed out that the magnitude of the non-monotonicity at $B \gtrsim B_c$, measured by the difference between the value of the $dV/dI$ in the zero current limit and minimum $dV/dI$ at $I_s$, never exceeds ~ 2% of the $dV/dI$ value at $I \to 0$. Therefore, qualitatively the same results are obtained if we plot $V/I$ vs $I$. If we analyze $dV/dI$ (or $V/I$) vs $V$, the characteristic voltages $V_s$ that correspond to $I_s$ in each $I$-$V$ curve, exhibit linearly increasing behavior with increasing $B$ and is almost independent of $T$.

To summarize, we have reported non-monotonic $dV/dI$ of the $B$-induced insulating films near the metal-insulator boundary. The observation may be evidence that the insulating state near $B_c$ consists of metallic domains connected by point contacts. In this picture, the non-monotonic $dV/dI$ arises as a consequence of the contrasting nonlinear transport properties of the metallic ($d^2V/dI^2 > 0$) and insulating phase ($d^2V/dI^2 < 0$). This interpretation implies that spatial inhomogeneity may play a dominant role in determining the nature of the $B$-induced metal-insulator transition occurring in the zero temperature limit.

The authors acknowledge V. Galitski for fruitful discussions. This work is supported by the NSF through Grant Nos. DMR-0239450 and DMR-0315320.